# Cycle expansions with pruned orbits have branch points

Ronnie Mainieri

*Complex Systems Group, MS B213,*
*Los Alamos National Laboratory, Los Alamos, NM 87545*

## Abstract

Cycle expansions are an efficient scheme for computing the properties of chaotic systems. When enumerating the orbits for a cycle expansion not all orbits that one would expect at first are present — some are pruned. This pruning leads to convergence difficulties when computing properties of chaotic systems. In numerical schemes, I show that pruning reduces the number of reliable eigenvalues when diagonalizing quantum mechanical operators, and that pruning slows down the convergence rate of cycle expansion calculations. I then exactly solve a diffusion model that displays chaos and show that its cycle expansion develops a branch point.

# Introduction

Cycle expansions are a very efficient way to compute the properties of chaotic systems. The range of chaotic systems to which they apply goes beyond dynamical systems. It includes spin systems in statistical mechanics [1] and few-body quantum systems [2]. It applies to these systems because underlying their dynamics there is a chaotic system. In the case of quantum mechanics the dynamical system is its classical approximation; in the case of statistical mechanics it is the (chaotic) dynamics that generates all configurations. This close relationship between classical dynamics, statistical mechanics, and quantum mechanics allows one to use the techniques of one area to solve the problems of another area. It also shows that the difficulties in one area should manifest themselves in another.

One of the unsolved problems in cycle expansions is the pruning of orbits. To compute a cycle expansion on has to enumerate all periodic orbits of the system [3, 4]. This is done by associating a code to each topologically possible orbit, and then finding the actual orbit based on the code. The code is made up of a few symbols — + and −, or 0 and 1 — but in general not all combinations of the symbols correspond to actual orbits of the dynamical system. When this occurs, one says that there is pruning of orbits. (The name comes from the representation of all the actual codes as a tree of symbols. A certain orbit not occuring corresponds to pruning a branch of this tree.) The problem that pruning introduces is that it slows down the convergence of cycle expansions.

The slow convergence in the presence of pruning is *not* a problem specific to cycle expansions. The problem is apparent in cycle expansions because other methods for computing quantities from chaotic systems are just not accurate or efficient enough [5, 6]. The problem of slow convergence is also not restricted to dynamical systems. It has been seen in quantum systems when the eigenvalues of are computed by diagonalizing an operator. If one wants to compute the n-th eigenvalue to a fixed precision, then the number of computer operations seems to be proportional to some power of $e^n$ [7]. The exponential character comes from the size of the matrix having to grow exponentially to attain the required accuracy. The convergence rate is largely affected by the number of periodic orbits that are pruned.

There is very little understanding of the effects of pruning and even less on how to control them. To gain a better understanding of the problem I have exactly evaluated the cycle expansion of a dynamical system with pruning. The interesting result is that the cycle expansion develops a branch point singularity. If there were no pruning the cycle expansion would be an entire function [8]. If one knew the nature of the singularity and how it develops, it would be possible to remove its effects from the cycle expansion. In the appendix I speculate that an algebraic branch point is the common case. If the type $x^{-p/q}$ of the this branch could be determined analytically, then it would be possible to numerically accelerate the convergence of cycle expansions or even to remove the singularity analytically.

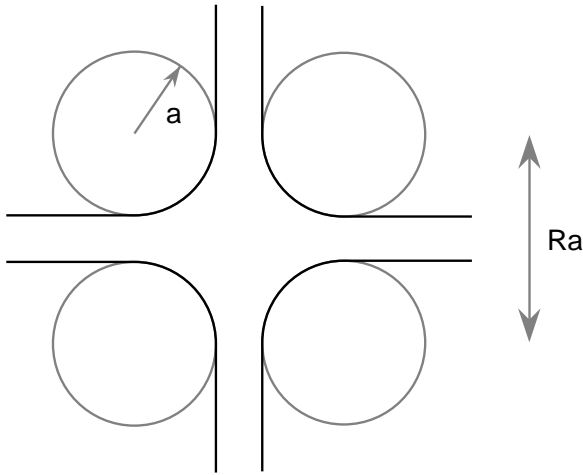

*Figure 1: Geometry of the scattering region. Wave packets enter from one of the channels and are then scattered in all directions by the hard walls. If, instead, a classical point particle came bumping down a channel, its motion would be chaotic.*

The model I use corresponds to a group of particles diffusing along the real line. The dynamics of each particle is independent of the others. I then ask the question: how does the area occupied by particles near the origin decay? The symbolic dynamics (codes for the orbits) of this problem is simple to determine and is heavily pruned. Despite the pruning, the number of orbits grows exponentially with the length of the orbit.

# Quantum resonances

To put the problem of pruning in perspective, I would like to show how it correlates to difficulties in doing quantum mechanical calculations with traditional numerical techniques. I will consider the computation of the resonances of a scattering problem that arises in a nano-junction (two nanometer scale wires joined at a point). This model has been treated by Roukes and collaborators [9, 10].

Suppose we want to solve the time independent Schrödinger equation in the infinite region delimited by four walls, as in figure 1. Each of the four corners is a quarter section of a circle. A particle contained within the walls cannot penetrate them (infinite potential beyond the walls). The problem is determining for what values of the energy $E$ the Schrödinger equation has a solution. After rescaling by the mass and Planck's constant we have to solve

$$\left(\nabla^2 + k^2(E)\right)\psi = 0,  \qquad (1)$$

subject to the boundary condition that $\psi$ is zero at the walls. In solving this equation we will obtain complex values for E, which are interpreted as being resonances of the system rather than bound states.

To solve this problem numerically one takes advantage of the zero potential inside the billiard and transforms it to a problem along the boundary of the billiard with the use of a Green function. This transforms a partial differential equation into an integral equation. A basis is then chosen to expand the integral equation along the boundary. A solution will only exist if a certain determinant is zero, the Korringa-Kohn-Rostoker determinant. The values of the energy for which the determinant is zero are the resonances. This method was first used for the study of chaotic systems by Berry [11].

In two dimensions, the Green function for the Schrödinger equation (equation 1) is given in terms of a Hankel function of the first kind of order zero

$$G(\mathbf{r}, \mathbf{r}') = -\frac{i}{4} H_0^{(1)}(k|\mathbf{r} - \mathbf{r}'|) \ . \tag{2}$$

If we choose a point $\mathbf{r}$ outside the billiard, then we have the identity

$$0 = \int d^2 r' (\nabla_{\mathbf{r}'}^2 + k^2) G(\mathbf{r}, \mathbf{r}') \psi(\mathbf{r}) - (\nabla_{\mathbf{r}'}^2 + k^2) G(\mathbf{r}, \mathbf{r}') \psi(\mathbf{r}') \ , \tag{3}$$

which can be transformed to an integral equation along the boundary by using the Green identity (integration by parts).

The same resonances may be computed using cycle expansions. The cycle expansion calculation is a semiclassical calculation based on a saddle point expansion of the path-integral representation of the wave-function. Semiclassical expansions of path integrals were pioneered by Morette [12]. The subject was greatly forward by a series of papers by Gutzwiller [13, 14], who determined the role of focal points along trajectories in semiclassical methods. But Gutzwiller's results are plagued with divergence problems. These where solved by Cvitanović and collaborators by using cycle expansions [2, 15].

In evaluating a cycle expansion only the periodic orbits of the system are considered. The expansion only depends on the set P of orbits that do not repeat themselves. For each orbit $\sigma$ one can compute its stability $\Lambda_\sigma$ as the largest eigenvalue of the monodromy matrix, and the Maslov index $\nu_\sigma$ which counts how many times the monodromy matrix looses rank as the orbit is traversed. The classical action of the orbit is given by $S_\sigma(\epsilon)$, which depends on the energy $\epsilon$ of the particle. The resonances of the system in two dimensions are given by the zeros of the semiclassical quantum Fredholm determinant [16]:

$$d(\epsilon) = \prod_{\sigma \in P} \prod_{k \geq 0} \left(1 - \frac{e^{iS_\sigma(\epsilon)/\hbar + i\pi\nu_\sigma/2}}{|\Lambda_\sigma|\Lambda_\sigma^k}\right)^{k+1} \ . \tag{4}$$

The enumeration of the orbits for the geometry of figure 1 is accomplished by noticing that a particle cannot collide with the same circular corner twice in a row. If the corners have a small radius of curvature (R is much bigger than a in figure 1), then the corners will appear as points, and very small changes in the trajectory of the particle will permit it to go from one corner to any other. If we number the corners 1 through 4, then the orbits can be enumerated by considering all sequences of those digits. The only restriction is that the same digit should not occur twice in a row. The first few orbits, and their codes, are illustrated in figure 2.

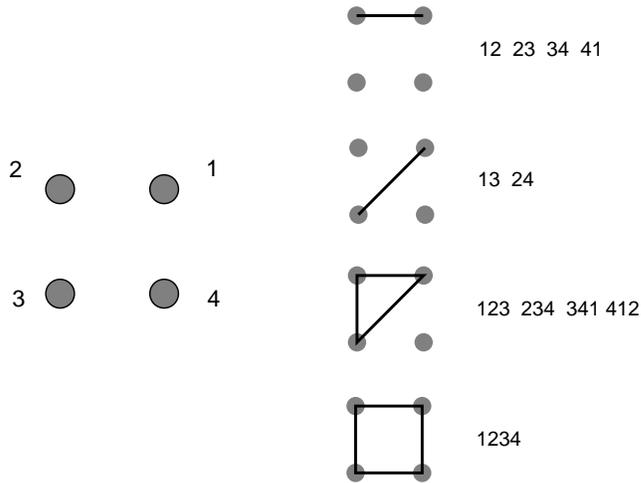

Figure 2: *For the semiclassical calculation all periodic orbits must be enumerated. If the disks are numbered 1 through 4 the periodic orbits may be found by searching for orbits that have a given disk-visitation sequence.*

The enumeration using four digits in the code is not the most efficient. There is a fourfold symmetry in the problem that could be used to reduce the number of orbits being counted and the size of the code necessary to specify each orbit. Also the formula for computing the resonances semiclassically, equation (4), can be improved. As written, the formula is a re-write of Gutzwiller's trace formula in product form. It is written that way to improve its convergence and to take advantage of shadowing of chaotic orbits [2]. Unlike other cycle expansions it does not come from the determinant of an operator. Recently Cvitanović and Vattay have managed to re-write it as a determinant [15].

The semiclassical calculation is very accurate for the lowest lying resonances. On a plot they are not distinguishable from the full quantum mechanical calculation [2, 17].

## Pruning of orbits

To compute the resonance using the full quantum method requires computing the eigenvalues of a matrix. As we vary the separation R between the disks (see

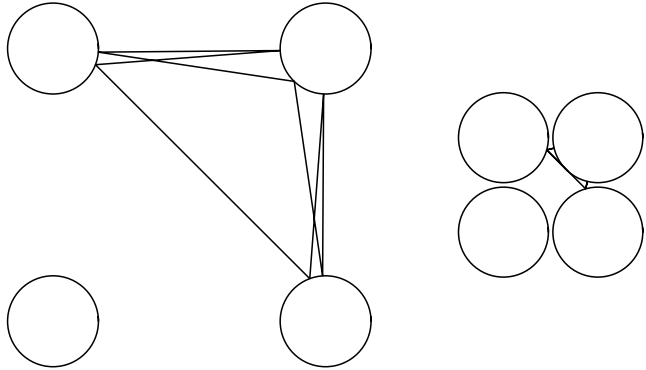

Figure 3: *The same orbit with two different inter-disk separations. As the disks get closer to each other, one of the disks may become tangent, and eventually intercept, a segment of a periodic orbit.*

figure 1), something remarkable happens. If the separation is large, R = 6a, then a small matrix, 14 × 14, is enough for the first five eigenvalues. But if the separation is small, R = 2.02a, then the matrix has to be much larger, 723×723, to achieve the same accuracy for the eigenvalues.

Is there something in the classical system that can be correlated with this difficulty? As the separation between the corners gets small, not all orbits predicted by the code of digits between 1 and 4 exist. Some of them disappear. The exact mechanism by which they disappear has been discussed by Hansen [18, 19]. In our case what happens is that as the corners approach each other, the segments of certain orbits approach the corners of the billiard (there is just less space for the orbit). Eventually the orbit becomes tangent to one of the corners. If the distance is reduced just a little more, then for the orbit to remain with the same code it would have to go through the corner it was tangent to (see figure 3). As this is not possible, the orbit no longer exists — it has been pruned.

The pruning of orbits slows down the convergence of cycle expansions. The slow convergence in the full diagonalization and in the cycle expansion seem to be correlated. The pruning itself has not been linked to the convergence of the quantum calculation, but the correlation is suggestive that the chaotic behavior of the underlying classical system is affecting the convergence of the quantum calculation.

## An exactly solved model

To better understand the effects of pruning I am going to introduce a simple model where the orbits are heavily pruned, which nevertheless can be solved exactly. The model is a point particle moving on the real line. For simplicity, it takes discrete steps and at each step it is in one of the intervals $[n, n+1)$ delimited by integers. This divides the real line into cells of length 1, as shown in figure 4. The dynamics is given by the map F, a map from the real line to itself. This map is constructed from another map, f, which is a map of the unit interval to itself. The map f is a two branch map and the point $1/2$ separates its right and left branches. At integer time steps the particle is displaced from

$$x_t \overset{F}{\mapsto} f(x_t \bmod 1) + \begin{cases} 1 & \text{if } x_t \bmod 1 < 1/2 \\ -1 & \text{if } x_t \bmod 1 > 1/2 \end{cases}. \qquad (5)$$

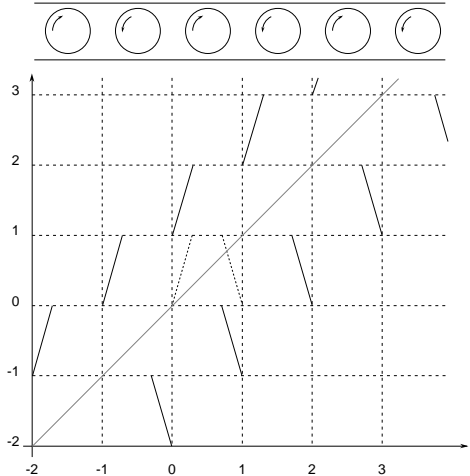

Figure 4: *The dynamics is given by a particle moving among many cells. At each time step the particle can go the neighboring left or right cell. It is as if there were a particle flowing among the convecting rolls in Rayleigh-Bénard convection.*

The effect of the global map F in equation (5) is that at after each time step the particle has moved to a different cell: either the right or the left neighboring cell. The map f may have a gap containing the point $1/2$. If a point falls in this gap it has then escaped the system and no longer participates in the dynamics. As time progresses, almost all points escape.

The symbolic dynamics of the map F is given by the symbolic dynamics of the map f. The map f is a two-branch one-dimensional map. All its trajectories can be encoded by using two symbols: $+$ for the right branch and $-$ for the left branch [20]. In the global map F, if the particle is mapped by the left branch of the map f, then it has also moved to the left neighboring cell. And if the particle is mapped by the right branch of the map f, then it has moved to the right. For the global map F the $+$'s and $-$'s now represent the sequence of left and right moves. For example, if a certain point x has a symbolic sequence starting with $++-\cdots$, that means that under the map F the point x has moved right, right,

Random walk solution    and then left.

Suppose that the origin-cell is uniformly covered with colored particles. As time goes by the particles diffuse through the system and the area occupied by the particles at the origin-cell decays. The colored particles that are at the origin-cell at time t are those particles that started there (it was the only cell with particles), wandered through the system, and returned to the origin-cell. These particles have taken right steps as often as they have taken left steps. In terms of the symbolic sequence of every orbit, the number of +'s in its code must be equal to the number of −'s in its code.

The initial set of particles occupies the unit interval $[0,1] = \Delta^{(0)}$. After one step some of the colored particles have left the system (trapped in the roll), some have moved to the right and occupy a segment in the cell $[1,2]$ and the some have moved to the left and occupy a segment in cell $[-1,0]$. As all the points in the segment that moved to the right remain connected, we can denote this smaller segment by $\Delta_+^{(1)}$. In the same way, the points that moved to the left are in segment $\Delta_-^{(1)}$. In this way, to every point x that has a certain symbolic sequence $\sigma = \sigma_0 \sigma_1 \sigma_2 \cdots$, there is a segment $\Delta_{\sigma_0 \sigma_1 \sigma_2 \cdots}^{(t)}$ with the same symbolic sequence which contains that point x.

If we denote the set of all orbits $\sigma$ that return to the origin in t steps by $R_t$ then the total length $Z_t$ occupied by the colored particles at the origin-cell is

$$Z_t = \sum_{\sigma \in R_t} |\Delta_\sigma^{(t)}| \ . \tag{6}$$

To solve the model exactly I will assume that the map f is of constant slope (see figure 7). Because of the uniform contraction it is simple to estimate the size $|\Delta_\sigma^{(t)}|$ of any segment from the contraction rate $1/\Lambda$:

$$|\Delta_\sigma^{(t)}| = \frac{1}{\Lambda^t} \ . \tag{7}$$

To compute the sum $Z_t$ we need to know how many segments there are at time t. That number is given by the number of ways one can go to the right or to the left and be at the origin cell at time t; this can only happen if t is even, as half the steps have to be to the right and half to the left. The number of different ways to arrange $t/2$ +'s and $t/2$ −'s is

$$\frac{t!}{(t/2)!(t/2)!} \ . \tag{8}$$

Using the Stirling asymptotic approximation for the factorial, one concludes that asymptotically

$$Z_t \to \left(\frac{2}{\Lambda}\right)^t \sqrt{\frac{2}{\pi t}} \ [\text{t even}] \ . \tag{9}$$

If $\Lambda > 2$, then $Z_t$ decays exponentially with t at a rate of $\gamma = \ln 2/\Lambda$ and should be simple to compute with cycle expansions. If the map where a tent map (and therefore $\Lambda = 2$) the behavior of $Z_t$ would be diffusive — $1/\sqrt{t}$ — rather than exponential.

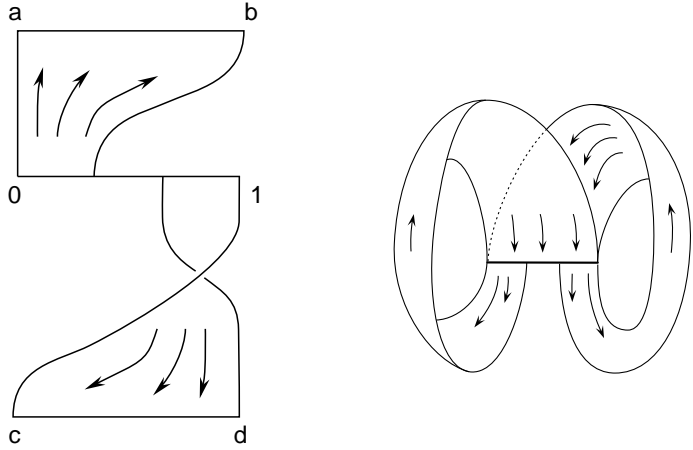

*Figure 5: The flow is formed by pasting together maps from one cell to its two neighboring cells. In the extended case point a is attached to 1 and point b to 2 and c and d are attached to −1 and 0. For the compact manifold a and d are attached to 0 and b and c to 1. The reduce flow attaches a $e^{\pm i\theta}$ to each flow so that one can keep track of the cell.*

Cycle expansion solution

We now want to determine the exponential rate of decay using cycle expansions. Rather than tracking the trajectories through the real line we will considered a reduced manifold. As each cell is equivalent to all others up to translations, we can keep track of which cell a point is in by carrying two numbers: the position $x$ within a unit cell and the cell number $n$. Neither $x$ nor $n$ are complex numbers, so they can be combined into one complex number

$$x_t \mapsto x_t e^{in\theta} \qquad (10)$$

from which it is still be possible to recover each of them. Each orbit point now carries an auxiliary field where $n$ is an integer and $\theta$ a dummy variable. Setting it to zero takes us back to the original dynamics. This operation is equivalent to compactifying the manifold where the dynamics takes place. Rather than a point flowing to a neighboring cell, it flows back to the same cell with the $e^{in\theta}$ part changing. This is illustrated in figure 5. In the compact manifold the dynamics is given by $G$,

$$x_t e^{in\theta} \stackrel{G}{\mapsto} \begin{cases} f(x_t) e^{i(n-1)\theta} & \text{if } x_t \bmod 1 < 1/2 \\ f(x_t) e^{i(n+1)\theta} & \text{if } x_t \bmod 1 > 1/2 \end{cases}. \qquad (11)$$

We can now repeat the computation for the total length $Z_t$ in the original cell occupied at time $t$ by the colored particles. We begin by considering a

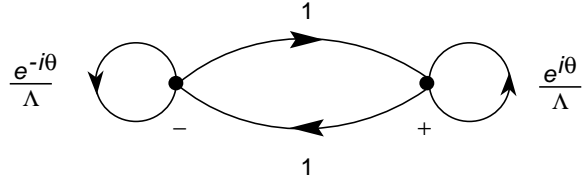

*Figure 6: Markov diagram for the stabilities of the map in the reduced manifold. A transition through the right branch of the map (+) gains a factor of $e^{i\theta}$, while a transition through the left branch (−) gains a factor of $e^{i\theta}$.*

generating function for the quantities $Z_t$, the zeta function

$$\zeta(z) = \exp\left(\sum_{n>0} \frac{z^n}{n} Z_n\right) . \tag{12}$$

The zero of $1/\zeta$ closest to the origin is related to the rate of decay. If $Z_n \to e^{\gamma n}$ then the term in the exponential sums to $-\ln(1 - ze^{\gamma})$ and $\zeta^{-1}$ has a zero at $e^{-\gamma}$.

To compute $Z_t$ in the reduced (compact) manifold we no longer need to worry about orbits returning to the origin. Suppose we computed the stability of all periodic orbits in the reduced manifold. In the extended manifold (the real line) some of these orbits would correspond to orbits that return to the origin and some would not. To each of these orbits we can associate a stability (product of the slopes along the orbit). The stability is computed by multiplying the slopes: $\Lambda e^{-i\theta}$ for the left branch and $\Lambda e^{+i\theta}$ for the right branch. If this procedure is adopted, the orbits that go through the left branch as often as they go through the right branch will not have a term of the form $e^{in\theta}$. This factor will cancel out.

I will now evaluate the total length $Z_t$ occupied by the colored particles at time t without the constraint that the orbits have to return to the origin-cell. Just as in equation (6) the length is given by

$$Z_t(\theta) = \sum_{\sigma \in \{+,-\}^t} |\Delta_\sigma^{(t)}| = \sum_{\sigma \in R_t} \frac{e^{in_\sigma \theta}}{\Lambda^t} , \tag{13}$$

where $n_\sigma$ is the cell number being occupied by the orbit in the extended space. It is obtained directly from the dynamics G in the reduced manifold. To restrict ourselves to only the orbits that return to the origin cell one can integrate in $\theta$ along the interval from 0 to $2\pi$. The integral of $e^{in_\sigma \theta}$ (with $n_\sigma \neq 0$) in that interval is zero, so all terms that correspond to orbits that do not return to the origin cell will be cancelled. We have then the relation

$$Z_t = \int_0^{2\pi} \frac{d\theta}{2\pi} Z_t(\theta) . \tag{14}$$

There is an advantage in first computing the sum $Z_t$ through $Z_t(\theta)$ rather than doing it directly. Because $Z_t(\theta)$ has no constraints, it can be computed by summing all the walks on the Markov diagram of figure 6. This is done by considering the matrix

$$M(\theta) = \frac{1}{\Lambda} \begin{bmatrix} e^{i\theta} & 1 \\ 1 & e^{-i\theta} \end{bmatrix} \qquad (15)$$

and noticing that

$$Z_t(\theta) = \mathrm{tr} M^t . \qquad (16)$$

We can combine these results on how to compute $Z_t$ and evaluate the zeta function in equation (12).

$$
\begin{aligned}
\zeta^{-1}(z) &= \exp\left(-\int \frac{d\theta}{2\pi} \sum_{n>0} \frac{z^n}{n} \mathrm{tr} M^n \right) \\
&= \exp\left(\int \frac{d\theta}{2\pi} \mathrm{tr} \ln(1 - zM) \right) \\
&= \exp\left(\int \frac{d\theta}{2\pi} \ln \det(1 - zM) \right) \qquad (17)\\
&= \exp\left(\int \frac{d\theta}{2\pi} \ln(1 - 2\frac{z}{\Lambda} \cos\theta) \right) . \qquad (18)
\end{aligned}
$$

The last of these expressions may be evaluated analytically, which is always satisfying in a problem that involves pruning of orbits. A method for carrying out integrals of this type is discussed in the Appendix. The result obtained in equation (17) will hold for more complicated maps, as will be derived later on. The result of the integral is

$$\zeta^{-1}(z) = \frac{1}{2}\left(1 + \sqrt{1 - \frac{4}{\Lambda^2} z^2}\right) . \qquad (19)$$

This function has no zeros, but it has a branch point at $z = \Lambda/2$, exactly at $e^{-\gamma}$, where $\gamma$ is the rate of decay. At the branch point the series expansion for $\zeta^{-1}$ around zero fails to converge.

Four-slope case  The same calculation can be carried out for a map that does not have constant slope. (The general case is discussed in the appendix.) One can use the method described there to compute the cycle expansion for a four-slope map, as in figure 7. For this case it is still possible to carry out the calculation analytically and one obtains a factorized cycle expansion written in terms of the stabilities $\Lambda_0$, $\Lambda_1$, and $\Lambda_{01}$ of the orbits on the reduced map

$$\zeta^{-1}(z) = \frac{1}{2}\left(1 - z^2\left(\frac{1}{\Lambda_{01}} - \frac{1}{\Lambda_0}\frac{1}{\Lambda_1}\right)\right)\left(1 + \sqrt{1 - a^2 - b^2}\right) , \qquad (20)$$

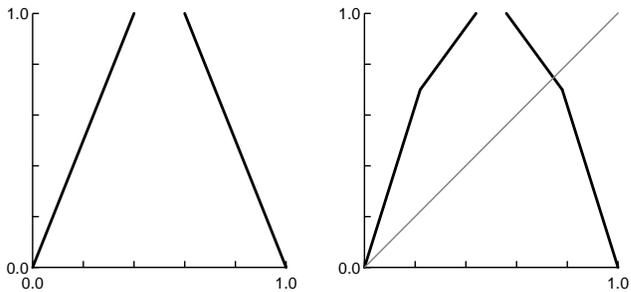

Figure 7: *The reduced map f with two slopes (left) or with four different slopes (right). This map should be used at the origin-cell in figure 4, and controls the "interesting" part of the dynamics. The general case of a two-branch map is discussed in the appendix.*

where $a$ and $b$ are also dependent on the slopes:

$$a = \frac{z(1/\Lambda_1 + 1/\Lambda_0)}{1 - z^2(1/\Lambda_{01} - 1/(\Lambda_0\Lambda_1))} \qquad (21)$$

and $b$ differs from $a$ by an $i$ and a sign

$$b = \frac{iz(1/\Lambda_1 - 1/\Lambda_0)}{1 - z^2(1/\Lambda_{01} - 1/(\Lambda_0\Lambda_1))} \; . \qquad (22)$$

In all cases the cycle expansion is factorized into two parts. One of the parts contains an algebraic root and the other has a polynomial form. The square root is responsible for the branch point of the cycle expansion. In the appendix I explain why I feel that such factorization is typical.

# Conclusions

I have exactly evaluated the cycle expansion for a system that has a large number of orbits pruned. I find that the cycle expansion factorizes into two parts, one of which contains an algebraic singularity (a feature I conjecture to be general when there is pruning). As the number of orbits grows exponentially (as derived in equation 9), one cannot attribute the singularity to an unusual type of chaos. I have also tried to correlate the slow down in convergence of the cycle expansion with the slow down in convergence in eigenvalue calculations.

Next, I would like to examine other possibilities on how singularities develop. As a parameter in the system varies, an orbit that exits, suddenly disappears. One can imagine that it is this sudden transformation that leads to the existence of branch points in the cycle expansion, and that if the pruning were done smoothly the branch points would disappear. I do not think that is the case. There are two ways in which one can envision "smoothing the pruning." One is to replace the hard walls of the billiard by smooth potentials. This does not change the character of the pruning [19, page 269]. The other possibility is to view the computation of the cycle expansion as a computation of the free energy of a spin system, and then smooth the interaction.

One can see the connection between dynamical systems and statistical mechanics by observing how stabilities are computed. To compute the stability of a periodic orbit we multiply the slopes of the map along the periodic trajectory. One can think of the periodic orbit being extended to infinity, repeating itself over and over. This infinite orbit would have a code. If we consider the code as a configuration of an Ising-like system, and the logarithm of the stability as the energy of the configuration, then the logarithm of the slope at each point can be considered as the interaction energy of one symbol with an infinite number of others. In this language the pruning of an orbit corresponds to a configuration having infinite energy (hard core condition), and therefore not occuring in the Gibbs ensemble [21].

Smoothing the pruning of orbits corresponds to giving the pruned orbits a large, but not infinite energy. As the energy of a configuration to be pruned increases, so do the correlations between distant symbols. For example, the orbit $(-+)^n+$ has to be pruned, but one only discovers that after observing the entire orbit. This corresponds to correlations of infinite length and a sign of phase transitions. This is why I do not feel that smoothing would eliminate the problem of branch points in cycle expansions.

It could also be that the system that I considered is too specific, and that if a more general type of pruning had been considered it would not have led to the algebraic singularity. I do not think this is the case because the pruning I considered ocurs in general. It is contained as a subset of most pruning rules. To see this one must view pruning rules as a formal language (in the sense of computer languages). If only a simple pruning rule is considered, for example, no two −'s in a row, then the resulting symbolic codes can be described by a regular language. Regular languages correspond to finite-shift dynamical systems [22]. If we continue adding further pruning rules, and eventually an infinite number of them, it is not clear what the resulting language would be for the possible symbolic codes (little seems to be known about the limit of formal languages).

Lets conjecture that the resulting language is more complicated than a regular language. In this case the language satisfies the pumping lemma: there are substrings $a$, $b$, and $c$, such that all symbols of the form $a^n b c^n$ are part of the language. This lemma allows you to pump-up the end parts of the word until it is as long as you want. The model solved is of this class, as $+^n(-+)-^n$ is a valid orbit. As all languages more complicated than regular languages satisfy the pumping lemma, then all of them will have the factorization and a singularity, as developed for the simple model.

# Acknowledgments

I would like to thank the Center for Nonlinear Sciences and the DOE for their financial support while this work was carried out. I would also like to thank Predrag Cvitanović and Brosl Hasslacher for their insightful comments.

# Appendix

Here I discuss how to compute the cycle expansion for maps f more complicated than a two-slope map. When the map has many different slopes the zeta function becomes an expression of the form

$$\zeta^{-1}(z) = \exp\left(\frac{1}{2\pi}\int d\theta \, \ln\left(a + R(e^{i\theta})\right)\right) , \qquad (23)$$

where R(x) is the rational function P(x)/Q(x) with P and Q polynomials. Further, Q has no constant term, Q(0) = 0. Both a and R depend on z, but that is not explicitly indicated.

First we will concentrate on the integral

$$I = \int_0^{2\pi} d\theta \, \ln\left(a + R(e^{i\theta})\right) , \qquad (24)$$

that appears in the expression for the zeta function. Applying the operator

$$\int da \, \frac{d}{da} \qquad (25)$$

to the integral I does not change it except for an additive constant. This constant may be determined by choosing R(x) to have coefficients that make the integral easy to compute.

$$\begin{aligned}
I &= \text{constant} + \int da \, \frac{d}{da} \int_0^{2\pi} d\theta \, \ln\left(a + R(e^{i\theta})\right) & (26) \\
&= C + \int da \int_\odot dz \, \frac{1}{iz} \frac{1}{a + R(z)} . & (27)
\end{aligned}$$

With the change of variables $z = e^{i\theta}$ the innermost integral is over the unit circle $\odot$. This requires us to assume the $R(z) + a$ has no poles on the unit circle. The constant is being called C.

$$I = C + \int da \, \Big( \sum_{\substack{z \\ a + R(z) = 0}} 2\pi \text{Res}\left(\frac{1}{z\,(a + R(z))}\right) + \frac{2\pi}{a} \Big) . \qquad (28)$$

There is a pole at $z = 0$ which can be easily computed, resulting in the $2\pi/a$ term. When it is integrated it results in a $2\pi \ln a$, which will result in the regular factor of the zeta function. One has

$$\zeta^{-1}(z) = e^{C/2\pi} a(z) \exp\left( \int da \sum_z \operatorname{Res}\left( \frac{1}{z(a+R(z))} \right) \right) \qquad (29)$$

The zeros of $a + R(z)$ cannot be computed explicitly except in a few simple cases (such as the two- or four-slope examples). The zeros are found by solving the polynomial equation

$$aQ(z) + P(z) = 0 , \qquad (30)$$

where the ratio of P and Q forms R. To carry out the integration $\int da$ one would have to know the functional relationship of the zeros with $a$. This relation is difficult to determine explicitly, but one does know that the zeros are algebraic functions in $a$. This means that the singularities of the zeta function (29) will be those that result from exponentiating the integral of an algebraic function. The exponential does not modify the character of any branch point, so if the zeta function has any branch points, they will result from integrating algebraic functions.

## Four-slope map

To evaluate the cycle expansion for the four-slope map, one proceeds just as in the case of the one-slope map. One constructs a Markov diagram, determines its transfer matrix with the appropriate factors of $e^{\pm i\theta}$ to impose the return to the origin-cell constraint, and then uses this matrix in equation 17 to compute the cycle expansion. In this case the slopes are not all given by a single number $\Lambda$, but rather by four different numbers, the four slopes. If we read the slopes from left to right along the x-axis the inverse slopes are: $s_{00}$, $s_{01}$, $s_{10}$, and $s_{11}$. If the map is symmetric two of those numbers will differ from the other two just by a sign.

The matrix M to be used in computation of the cycle expansion in equation 17 is

$$\begin{bmatrix} s_{00} e^{-i\theta} & s_{01} e^{-i\theta} & 0 & 0 \\ 0 & 0 & s_{01} e^{i\theta} & s_{01} e^{i\theta} \\ 0 & 0 & s_{11} e^{i\theta} & s_{11} e^{i\theta} \\ s_{10} e^{-i\theta} & s_{10} e^{-i\theta} & 0 & 0 \end{bmatrix} \qquad (31)$$

From this matrix one can obtain the cycle expansion in equation 20.